\documentclass[prl,twocolumn,showpacs,showkeys]{revtex4}

\usepackage{graphics}
\usepackage{graphicx}

\newcommand{\be}{\begin{equation}}
\newcommand{\ee}[1]{\label{#1} \end{equation}}
\newcommand{\ba}{\begin{eqnarray}}
\newcommand{\ea}[1]{\label{#1} \end{eqnarray}}
\newcommand{\nl}{\nonumber \\}

\begin{document}

\title{{\bf Non-Extensive Black Hole Thermodynamics Estimate for Power-Law Particle Spectra}}

\author{ 
  T. S. Bir\'o
}

\affiliation{
  {KFKI Research Institute for Particle and Nuclear Physics,} \\
  { H-1525 Budapest, P.O.Box 49, Hungary} 
}

\keywords{non-extensive thermodynamics, heavy ion collisions, black hole evaporation}
\pacs{65.40.gd, 25.75.Ag, 04.70.Dy }

\date{\today}

\begin{abstract}
  We point out that by considering the Hawking-Bekenstein entropy of
  Schwarzschild black hole horizons as a non-extensive Tsallis entropy,
  its additive formal logarithm, coinciding with the Renyi entropy,
  generates an equation of state with positive heat capacity above a
  thereshold energy. 
  Based on this, the edge of stability is conjectured to be trans-Planckian,
  i.e. being in the quantum range. From this conjecture an estimate arises for the
  q-parameter in the Renyi entropy, ($q=2/\pi^2$), also manifested in the canonical power-law distribution of
  high energy particles ($q \approx 1.2$ for quark matter). 
\end{abstract}

\maketitle
\normalsize



\section{Introduction}

Non-extensive thermodynamics \cite{NEXT1} aims at describing dynamically and statistically entangled
systems, among others also the quark-gluon plasma at the moment of hadronization.
By using a non-Boltzmannian formula for the entropy, the experimentally observed power-law tailed
particle spectra can be interpreted as reflecting a canonical ensemble in the
non-extensive thermodynamics \cite{NEXT2,NEXT3,NEXT4,NEXT5}. 
The theoretical prediction of the power $\nu$, related to
the Tsallis parameter, $q=1+1/\nu$, is however a very difficult problem.

In an earlier work we have fitted several hadron spectra observed at
RHIC \cite{PHENIX1,PHENIX2,PHENIX3,PHENIX4,STAR1,STAR2,STAR3} 
by assuming a blast wave picture and quark recombination \cite{SQM2008}. 
These fits agreed with
a common Tsallis parameter of $q \approx 1.2$ for the quark matter quite remarkably.
According to the quark coalescence picture the corresponding value of $(q-1)$ for mesons
should be the half, and for baryons one third of this fit. Unfortunately the experimental
data are not yet precise enough to prove or disprove this model.
On the other hand, in \cite{EPL} it was shown that the cut power-law energy distribution
may arise from a repeated non-additive combination of small amounts of energy. Starting
with a general $Q^2=2E_1E_2(1-\cos\theta)$ - dependent correction to the addition of the respective
kinetic energies of massless partons, $E_1$ and $E_2$, i.e. considering a combination law
of the form
\be
 E_1 \oplus E_2 = E_1 + E_2 + G(Q^2)
\ee{E_COMBO}
in the large number limit and neglecting the angular dependence 
one arrives at a general asymptotic combination law of $E_1\oplus E_2=E_1+E_2+aE_1E_2$.
This combination law is equivalent to the addition of formal 
logarithms \cite{FORMLOG},
\be
 L(E) = \frac{1}{a} \ln \left( 1 + a E\right),
\ee{FORM_LOG}
leading to
\be
 L(E_1\oplus E_2) = L(E_1) + L(E_2).
\ee{LE_COMBO}
As a consequence the statistical distribution which arises  by using such non-additive
composition rules is
\be
f(E) = \frac{1}{Z} \, e^{-\beta L(E)} = \frac{1}{Z} \, \left(1+aE \right)^{-1/(aT)}
\ee{CAN_DIST}
Here $a=G'(0)$, the derivative of the correction function at low $Q^2$
is non-perturbative, so there is no easy theoretical caclulation to predict its value.

On the other hand such an energy-distribution can also be derived 
as a canonical distribution stemming from the Tsallis ($S_T$) or 
Renyi-entropy ($S_R$) \cite{RENYI1,RENYI2,GE1,GE2,GE3,GE4}.
Both entropy formulas contain a $q$ parameter and are in fact connected as
\be
S_R = \frac{1}{q-1} \ln \left(1+(q-1)S_T \right).
\ee{RENYI_TSALLIS}
The Renyi entropy is defined as
\be
 S_R = \frac{1}{q-1} \ln \sum_i  p_i^q,
\ee{RENYI_DEF}
and in the $q \rightarrow 1$ limit it coincides with Boltzmann's entropy formula.
The canonical energy distribution is derived from maximizing
\be
 S_R - \beta \sum_i p_iE_i  - \alpha \sum_i p_i.
\ee{SRENYI_MAX}
Differentiation with respect to $p_i$ leads to
\be
p_i = \frac{1}{Z} \left(1+\frac{\beta(E_i-\mu)}{1-q} \right)^{\frac{1}{q-1}}.
\ee{RCANON_F}
Here $\mu$ can be expressed by $\alpha$ and the other parameters, $Z$ is set acording
to the normalization condition $\sum_i p_i = 1$.

The main difference is that while the Renyi entropy is additive for factorizing probabilities,
the Tsallis entropy is not. The canonical distribution is the same for both entropies,
one being a monotonous function of the other, but their stability properties -- connected to
the second derivative -- differ as a rule.

In this paper we attempt to obtain an estimate for the $q$ parameter of the Renyi and
Tsallis entropy. Lining up with the conformal field theory -- gravity conjecture 
\cite{MALCADENA,WITTEN,KLEBA}
which gave already an estimate for the minimal viscosity of matter \cite{SON}, 
now we investigate
simple, static and radial Schwarzshild black hole thermodynamics and derive an estimate
for the minimal $q$ parameter which stabilizes their equation of state above an energy
considered to be in the quantum gravity (trans-Planckian) range. This is a two step
process: First we demonstrate that the interpretation of the Bekenstein-Hawking
entropy\cite{BH1,BH2,BH3} as a non-extensive Tsallis entropy and the use of the additive Renyi
entropy instead stabilizes the Schwarzschild black hole horizon equation of state
above a given energy. Then we ask the question whether this energy, the inflection
point of the $S(E)$ curve, is at or below the ground state energy of a semiclassical
string spanned over the diameter of the horizon, $2R=4E$ in Planck units.
All over this paper energy, mass and momentum is measured in multiples of the Planck mass, $M_P$,
while length and time in multiples of the Planck length, $L_P$. All equations relating
unlike quantities are to be supported by corresponding powers of $M_P$ and $L_P$.
The entropy is measured in units of the Boltzmann constant, $k_B$.
In a $c=1$, $k_B=1$ system one expresses the Planck constant as $\hbar = L_P M_P$ and 
Newton's gravity constant as $G=L_P / M_P$.

\section{Renyi entropy for black hole horizon}

The Bekenstein-Hawking entropy for simple (static, radial-symmetric) black hole horizons
can easily be obtained from Clausius' entropy formula,
\be
 S = \int \frac{dE}{T},
\ee{CLAUSIUS}
by using the Unruh temperature\cite{UNRUH} associated to the acceleration 
at the surface of the coordinate singularity (red shift factor corrected surface gravity
of the horizon):
\be
 T = \frac{g}{2\pi}.
\ee{UNRUH}
For the normal gravitational acceleration at Earth's surface this temperature is 
very small, $k_BT \approx 10^{-19}$ eV, but in a relativistic heavy ion collision,
assuming a stopping from the speed of light to zero in a distance equal to the half
Compton wavelength of a proton, meaning a deceleartion of $g=c^2/2\lambda=mc^3/\hbar$,
the Unruh temperature is in the range of $k_BT = mc^2/2\pi \approx 150$ MeV. 
Indeed particle spectra stemming from such collisions agree well with thermal model
estimates with temperatures of this magnitude. It has already been proposed
that such a deceleration could be the general reason behind observing thermal-like
spectra born in sudden, dramatic hadronization events\cite{KHARZEEV,SatzKharzeev}.

In general for a metric given by
\be
 d\tau^2 = f(r) dt^2 - \frac{dr^2}{f(r)^2} - r^2 d\Omega^2
\ee{METRIC}
with $t$ and $r$ being the time and radius coordinates for the far, static observer
and $d\Omega$ the two-dimensional surface angle, the acceleration of a test particle
with mass $m$ follows from the Maupertuis action
\be
I = -m \int d\tau.
\ee{MAUPERTUIS}
The radial equation of motion derived from the corresponding Lagrangian
\be
L = \sqrt{f\dot{t}^2-\dot{r}^2/f-\ldots},
\ee{LAGRANGIAN}
with $\dot{t}=dt/d\tau$ and $\dot{r}=dr/d\tau$ and supressing terms related to
angular components, is given by
\ba
f\dot{t} &=& K, \nl
f\dot{t}^2-\dot{r}^2/f &=& 1
\ea{EOM}
with an integral of motion related to the energy, $K$.
Eliminating $\dot{t}$ from the equations one obtains
\be
 \dot{r}^2=K^2-f(r),
\ee{RAD_EOM}
whose $\tau$-derivative delivers the radial acceleration
\be
 \ddot{r} = -\frac{1}{2} f'(r) = -g.
\ee{RAD_ACCEL}
The corresponding Unruh temperature becomes
\be
 T = \frac{1}{4\pi} f'(r).
\ee{RAD_UNRUH}
Considering now that the internal energy is practically the mass one obtains the entropy
\be
 S = 4\pi \int \frac{dM}{f'(r)}.
 \ee{BH_ENTROPY}
This result can be written in a more elegant form by noting that the denominator,
$f'(r)$ -- to be evaluated at the condition $f(r)=0$ -- is a Jacobian for a
Dirac-delta constraint. Therefore the above BH-entropy equals to
\be
 S = 4\pi \int\!\!\!\int \delta(f(r,M)) \, dr dM.
\ee{BH_SHELL}
This form reminds to a microcanonical shell in the phase space of the variables $r$ and $M=E$.

In the case of the Schwarzschild black hole solution to the Einstein equations one
has
\be
f(r) = 1 - \frac{2GM}{c^2r} = 1 - \frac{R}{r},
\ee{SCHWARZ}
The horizon condition $f(r)=0$ is fulfilled at $r=R=2M=2E$ in Planck units ($G=1$).
The acceleration is $g=1/2R=1/4E$ and the BH-entropy becomes
\be
 S = 4\pi \int \frac{R}{2} dR = \pi R^2.
\ee{AREA_LAW}
Since the same result emerges for any $f(r,M)=1-2GM/r-h(r)$, linear in $M$,
the entropy of such simple black hole horizons is proportional to their area.

\begin{figure}
\begin{center}
	\includegraphics[width=0.38\textwidth,angle=-90]{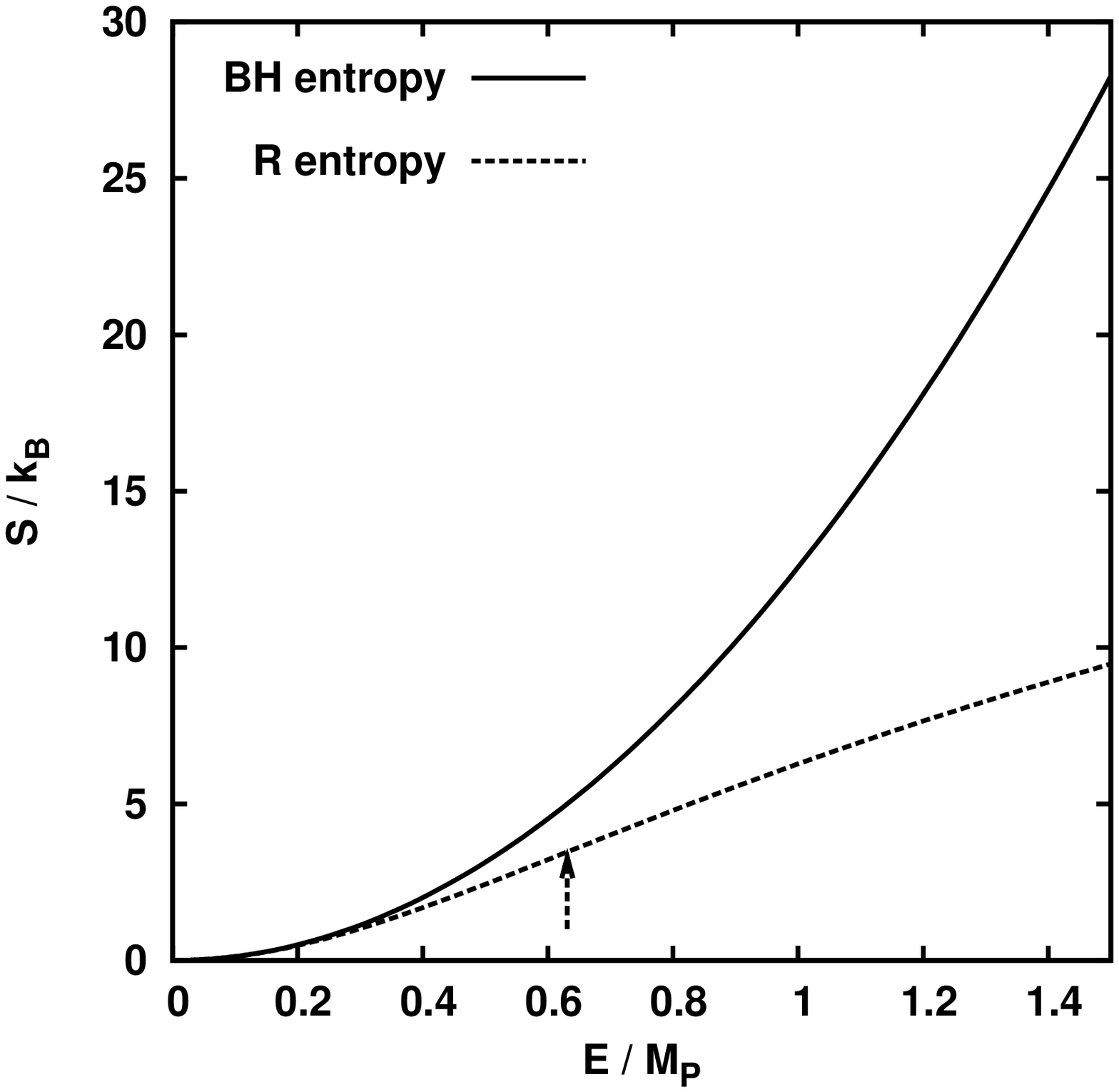}
	\includegraphics[width=0.38\textwidth,angle=-90]{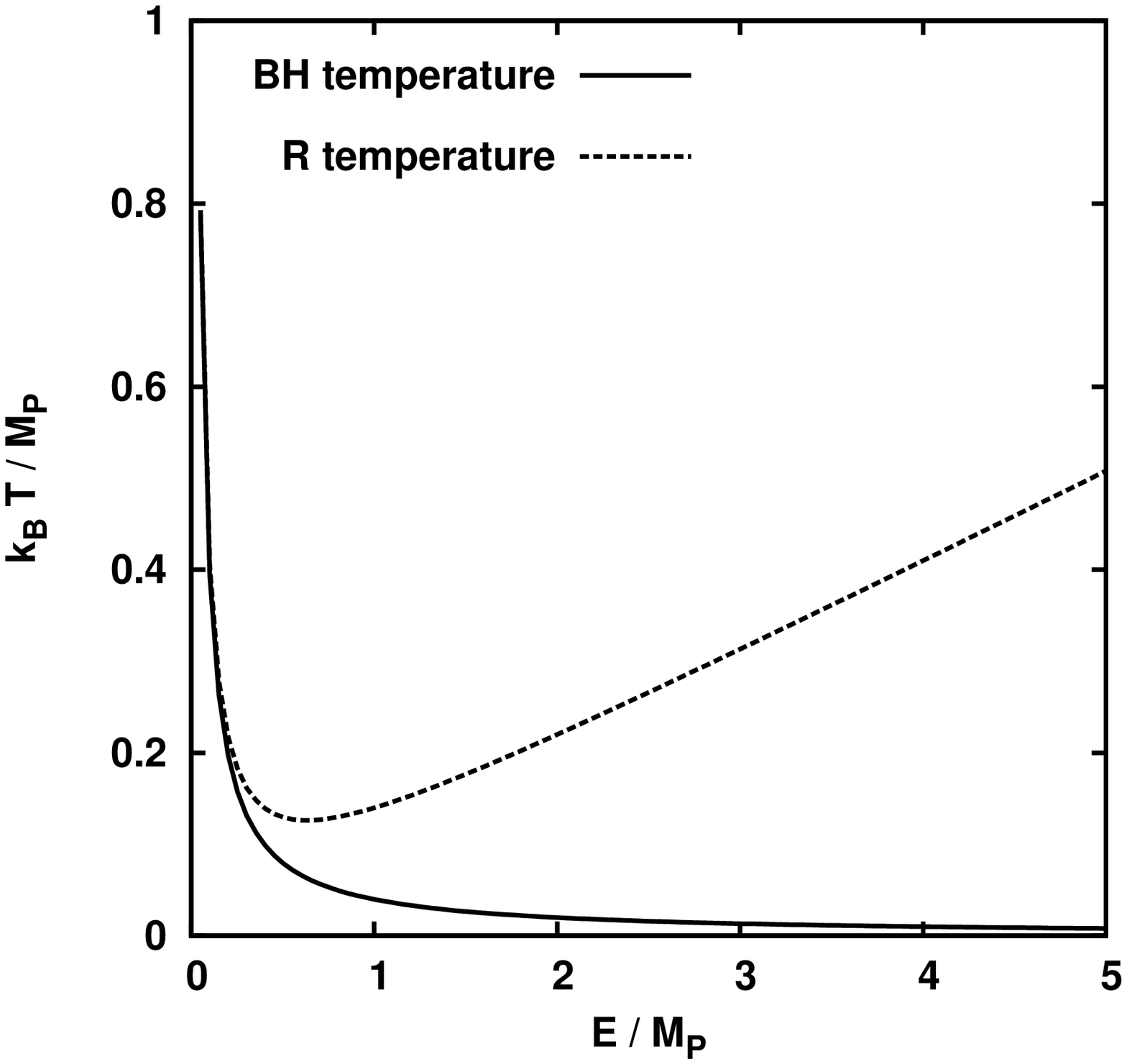}
	\includegraphics[width=0.38\textwidth,angle=-90]{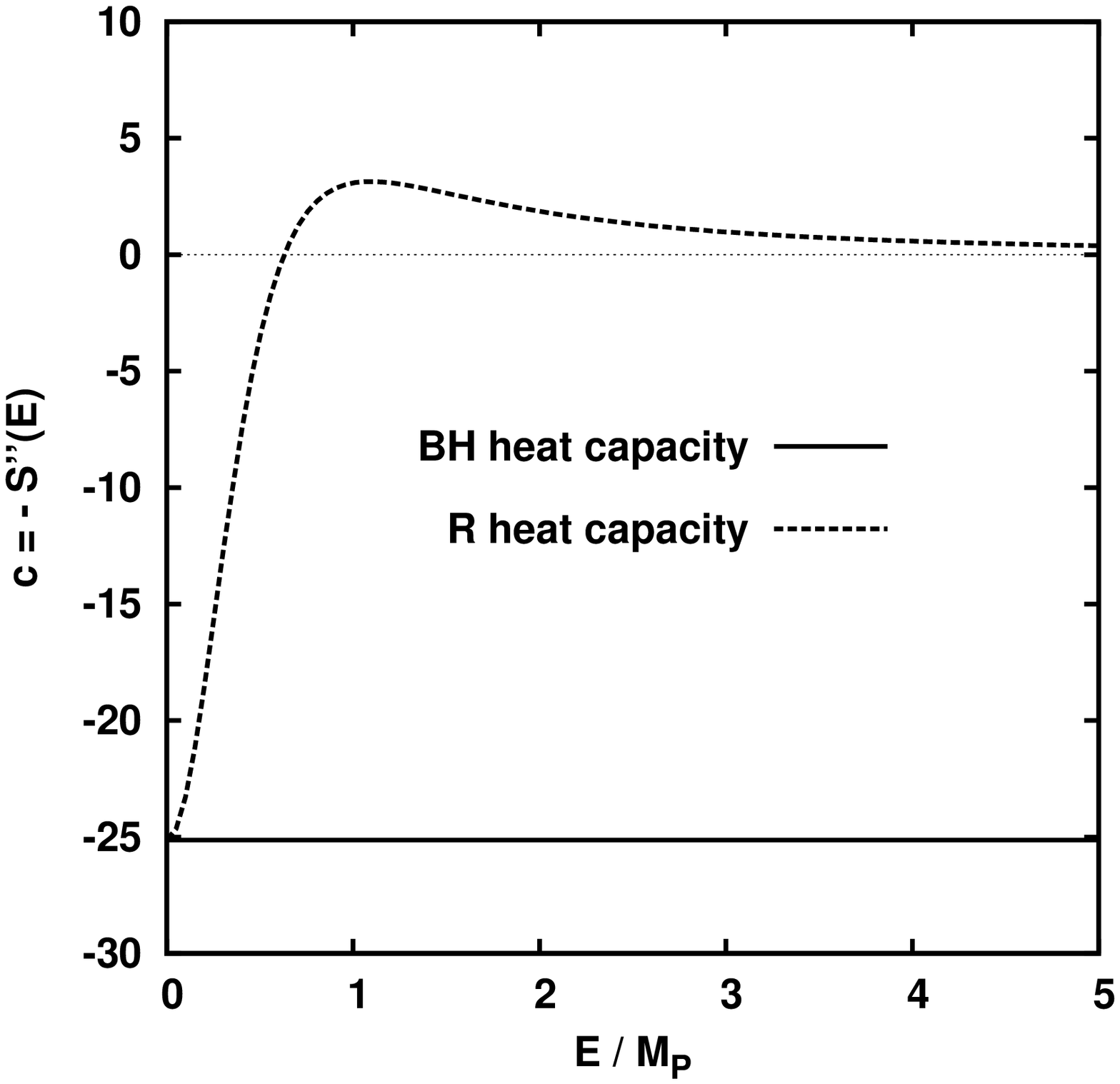}
\end{center}
\caption{ \label{Fig1}
  The Schwarzschild black hole horizon entropy, temperature and heat capacity
  as a function of the total internal energy (mass inside the horizon).
  The solid curves correspond to the Bekenstein-Hawking entropy,
  while the dashed curves to its formal logarithm, the Renyi entropy.
}

\end{figure}

This result, however, leads to an equation of state, $S(E)$ which describes an object
with negative heat capacity:
\ba
 S &=& 4\pi E^2, \nl
 \frac{1}{T} &=& S'(E) = 8\pi E, \nl
 c &=& - S''(E) = -8\pi.
 \ea{BH_EOS} 
Considering now the Hawking-Bekenstein entropy as a Tsallis entropy, since nobody knows
whether it were additive if two black holes would be united, one is tempted to use
its additive form, the Renyi entropy for defining the equation of state.
Using equation (\ref{RENYI_TSALLIS}), one obtains the following:
\ba
S &=& \frac{1}{a} \ln \left( 1 + 4\pi a E^2 \right), \nl
\frac{1}{T} &=& S'(E) = \frac{8\pi E}{1 + 4\pi a E^2 }, \nl
c &=& - S''(E) = 8\pi \frac{4\pi a E^2-1 }{\left(1 + 4\pi a E^2 \right)^2}
\ea{RENYI_EOS} 
with $a=q-1$.
Figure \ref{Fig1} shows the $S(E)$ (upper frame), the $T(E)$ and the $c(E)$ curves
for both the Bekenstein-Hawking and Renyi equation of state for Schwarzschild black holes.


It is straightforward to inspect that above the inflection point in energy,
\be
E_0 = \frac{1}{2\sqrt{a\pi}}
\ee{INFLECTION}
the heat capacity is positive and the black hole is a thermodynamically stable object.
Below this energy the heat capacity continues to be negative, describing an unstable object.
The temperature at this energy,
\be
T_0 = \frac{1}{8\pi E_0} + \frac{a}{2} E_0 = \frac{1}{2} \sqrt{\frac{a}{\pi}},
\ee{MIN_TEMP}
is finite, unless one considers the Boltzmann limit by $a \rightarrow 0$ ($q \rightarrow 1$).


\section{Semiclassical estimate for the black hole $q$ parameter}

Classical Schwarzschild black hole horizons would be thermodynamically stable if the edge of
stability, $E_0$, lied below the quantum mechanical ground state energy.
Of course, in the absence of a functioning quantum theory of gravity, one can only
have an estimate for this value from semiclassical considerations. The simplest of which
is regarding the energy of a string of the length of the diameter, $2R$, having a wave
number and frequency $\omega = k = \pi/2R$ (the sinus wave with no intermediate nodes).
The corresponding ground state energy is required to be greater than the inflection point
of the $S_R(E)$ curve:
\be
 \frac{\hbar\omega}{2} = \frac{\pi}{8E_0} \ge E_0.
\ee{GROUND_STATE}
Utilizing the relation (\ref{INFLECTION}) one concludes that it is equivalent to
the condition
\be
  a = q-1 \ge \frac{2}{\pi^2}.
\ee{qESTIMATE}
It is amazing to note that this estimate, $q\approx 1.2026$, how well approximated is
by cosmic ray observations ($q=11/9$) \cite{BECK1,BECK2} and by the quark coalescence fit to
RHIC hadron transverse momentum spectra ($q \approx 1.2$) \cite{SQM2008}.

In summary, by interpreting the Bekenstein-Hawking entropy as a non-extensive Tsallis
entropy of simple black hole horizons, and therefore considering their equation of state
based on the Renyi entropy, enables such horizons to be thermodynamically stable above
a given energy. Requiring that this energy belongs to a semiclassical ground state
of a string stretched over the diameter of the horizon, amazingly an estimate for the $q$-parameter
of Renyi's and Tsallis' entropy formulas arises. This estimate is close to findings in
relativistic heavy ion collisions and in cosmic ray observations. 
This result does not rely on the AdS -- CFT duality, 
as estimates for the viscosity to entropy density ratio did\cite{GYULASSY,BUCHEL1,BUCHEL2,HIDAKA}.

Of course, we do not suggest that 3-dimensional, gravitational (mini) black holes
would form in high energy particle and heavy ion collisions. However, due to
the extreme deceleration by the stopping, a Rindler horizon may occur for the newly
produced hadrons. This can be, in general, the origin of thermal looking spectra
as suggested earlier by Kharzeev\cite{KHARZEEV,SatzKharzeev}. This also can be the
mechanism behind the leading non-extensive effect in the power-law spectra, consistent
with a statistical, constituent quark matter hadronization picture.
It has been recently reported about ultrashort laser pulse experiments producing
among other known radiations also a Hawking radiation of photons\cite{LASER}.
The authors emphasize that no black hole formation is needed in the classical sense --
the formation of an event horizon suffices to produce features analogue to the
Hawking radiation in the spectrum.

\section*{Acknowledgement}

This work has been supproted by the Hungarian National Science Fund, 
OTKA (K68108).
Discussions with Prof. B. M\"uller and A.Jakov\'ac are gratefully acknowledged.

\end{document}